\documentclass[apj]{emulateapj}   
\usepackage{graphicx}
\usepackage{multirow}

\usepackage{amsmath}
\usepackage{natbib}
\usepackage{hyperref}
\usepackage[usenames,dvipsnames]{color}  

\usepackage[normalem]{ulem}

\def\msun{{\rm ~M}_{\odot}}
\def\rsun{{\rm ~R}_{\odot}}
\def\zsun{{\rm ~Z}_{\odot}}
\def\gpy{{\rm ~Gpc}^{-3} {\rm ~yr}^{-1}}
\def\kms{{\rm ~km} {\rm ~s}^{-1}}

\begin{document}

\title{The most ordinary formation of the most unusual double black hole merger}

\author{
   Krzysztof Belczynski\altaffilmark{1}
}

\affil{
   $^{1}$ Nicolaus Copernicus Astronomical Center, Polish Academy of Sciences,
          ul. Bartycka 18, 00-716 Warsaw, Poland (chrisbelczynski@gmail.com)
}

\begin{abstract}
LIGO/Virgo Collaboration reported the detection of the most massive black hole --- 
black hole (BH-BH) merger up to date with component masses of $85\msun$ and $66\msun$
(GW190521). Motivated by recent observations of massive stars in the 30 Doradus cluster 
in the Large Magellanic Cloud ($M_\star \gtrsim 200\msun$; e.g. R136a) and employing 
newly estimated uncertainties on pulsational pair-instability mass-loss (that 
allow for possibility of forming BHs with mass up to $M_{\rm BH} \sim 90\msun$)
we show that it is trivial to form such massive BH-BH mergers through the classical 
isolated binary evolution (with no assistance from either dynamical interactions
or exotica). 
A binary consisting of two massive ($180\msun$ + $150\msun$) Population II stars 
(metallicity: $Z\approx0.0001$) evolves through a stable Roche lobe overflow and common 
envelope episode. Both exposed stellar cores undergo direct core-collapse and form 
massive BHs while avoiding pair-instability pulsation mass-loss or total disruption. 
LIGO/Virgo observations show that the merger rate density of light BH-BH mergers 
(both components: $M_{\rm BH}<50\msun$) is of the order of $10-100\gpy$, while 
GW190521 indicates that the rate of heavier mergers is $0.02-0.43\gpy$. Our model (with 
standard assumptions about input physics) but extended to include $200\msun$ stars 
and allowing for the possibility of stellar cores collapsing to $90\msun$ BHs produces 
the following rates: $63\gpy$ for light BH-BH mergers and $0.04\gpy$ for heavy BH-BH 
mergers. We do not claim that GW190521 was formed by an isolated binary, but it 
appears that such a possibility can not be excluded.  
\end{abstract}

\keywords{stars: black holes, neutron stars, x-ray binaries}

\section{Introduction}
\label{sec.intro}

LIGO/Virgo Collaboration (LVC) has reported the discovery of a surprisingly heavy double 
black hole (BH-BH) merger with component masses $m_1=85^{+21}_{-14}\msun$ and 
$m_2=66^{+17}_{-18}\msun$ and an effective spin parameter $\chi_{\rm eff}=0.08^{+0.27}_{-0.36}$ 
at redshift $z=0.82$ (GW190521; ~\cite{gw190521a}). The corresponding merger rate density 
of events similar to GW190521 was estimated to be $0.13^{+0.30}_{-0.11}\gpy$.

Stars are not expected to form BHs of such masses. In particular, the Pair-instability Pulsation 
Supernovae (PPSN; \cite{Heger2002,Woosley2007}) are associated with severe mass loss that 
limits BH mass and Pair-instability Supernovae (PSN; \citep{Bond1984a,Fryer2001,
Chatzopoulos2012a}) are expected to completely disrupt massive stars with no resulting BH 
formation. These processes were believed to create the so called upper mass-gap in the BH mass 
spectrum i.e., the lack of stellar-origin BHs in the mass range $M_{\rm BH} \sim 50-135\msun$
~\cite{Marchant2016,Mandel2016a,Belczynski2016c,Spera2017}). It appeared that results of O1/O2 
advanced LIGO/Virgo observations were consistent with the existence of this mass gap
~\citep{Fishbach2020b}. Yet, the latest LIGO/Virgo O3 observations revealed GW190521. 

This has naturally promoted proposals in which BHs in GW190521 are not products of standard 
stellar evolution. These proposals include dynamical formation scenarios of repeated BH 
mergers in dense clusters~\citep{Rizzuto2020,Fragione2020,Gayathri2020}, repeated stellar 
mergers in dense clusters~\citep{DiCarlo2019,DiCarlo2020,Renzo2020b}, BH captures in
galactic nuclei~\citep{Gondan2020}, primordial black holes ~\citep{DeLuca2020}. Some more 
exotic scenarios are also being put forward such as head-on collisions of boson stars
~\citep{Calderon2020}. Alternatively, it is claimed that the LVC analysis is not the only 
solution to the GW190521 waveform and the actual BH masses may be outside the upper mass gap 
and are consistent with standard stellar evolution ~\citep{Fishbach2020c,Moffat2020,Nitz2020}. 

In the last few years the understanding of the upper mass gap begun to change. First, it was 
proposed that the first population of metal-free (Population III) stars may form BHs up to 
$\sim 70\msun$ without violating the pair-instability physics~\citep{Woosley2017}. This was 
extended to $\sim 85\msun$ by recent detailed stellar evolution~\citep{Farrell2020,Tanikawa2020} 
and population synthesis calculations~\citep{Kinugawa2020}. Second, it was proposed that for 
the intermediate-metallicity stars (Population II) BHs can form with masses up to $80\msun$
~\citep{Limongi2018}. 
Third, for high-metallicity stars (Population I) the limit was increased to $70\msun$
~\citep{Belczynski2020a}. These updates on position of lower edge of the upper mass gap 
were the result of detailed considerations of stellar evolution processes (e.g., rotation, 
mixing, convection) that allow some stars to avoid the PPSN/PSN. Finally, it was shown that
for low metallicity stars ($Z<10^{-5}-10^{-4}$), the uncertainties in the reaction rate of 
carbon burning (along with uncertainties on mixing/dredge-up) can potentially shift 
the onset of the BH upper mass gap up to $90\msun$~\citep{Farmer2020,Costa2020}. This reaction 
rate concerns one of the most uncertain reactions used in stellar evolution and yet it plays 
really important role in astrophysics~\citep{deBoer2017,Takahashi2018,Holt2019,Sukhbold2020}. 

Here, we adopt the latest results on the lower bound of the upper mass-gap to test whether 
it is possible to {\em (i)} form BH-BH mergers with masses as reported by LVC for GW190521 
and {\em (ii)} whether it is possible to form enough of them to match the LVC reported merger 
rate of such events. We perform our analysis in the framework of the most ordinary BH-BH 
merger formation scenario: the classical isolated binary evolution of Population I/II stars.

\begin{figure}
\hspace*{-0.4cm}
\includegraphics[width=0.5\textwidth]{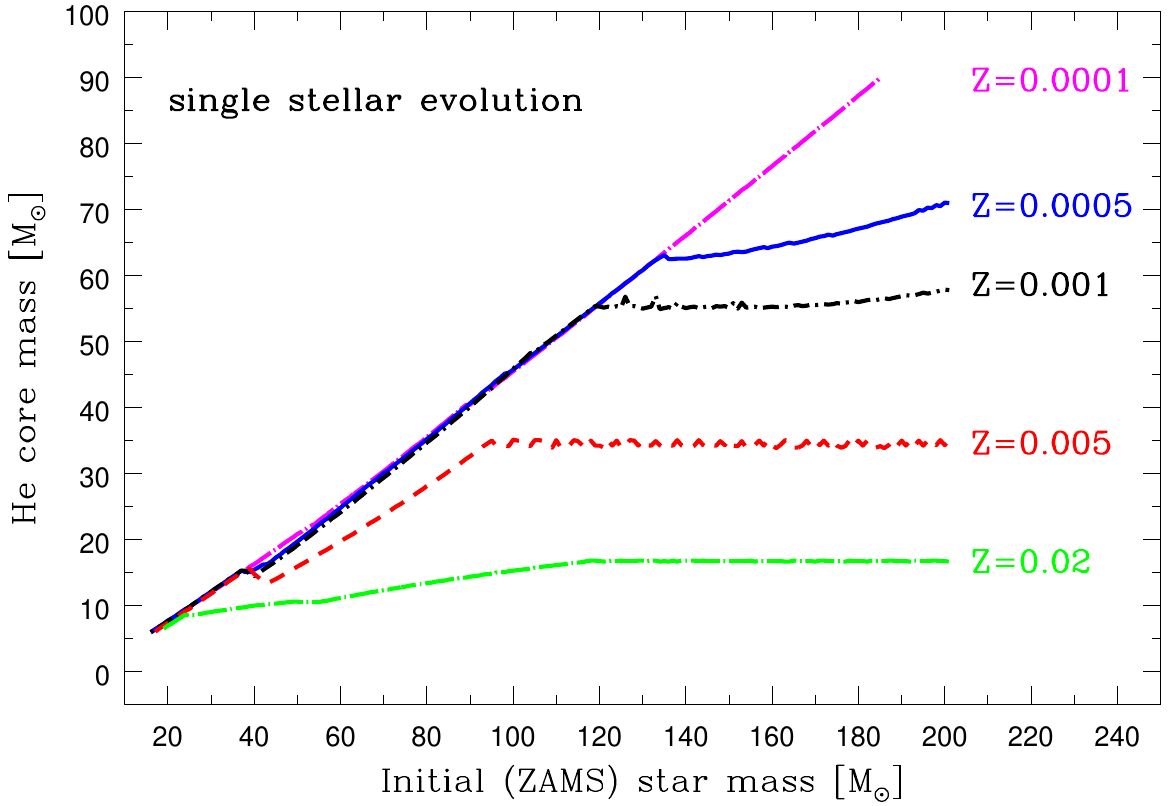}
\caption{
Initial star mass --- final helium core mass relation for single star 
evolution for various metallicities. Only stars that form black holes 
are shown. Helium core mass is a good approximation of the black hole 
mass especially for stars in close binaries that form BH-BH mergers as the 
binary interactions (RLOF, CE) remove H-rich stellar envelopes. Note 
that massive helium cores ($M_{\rm He} \gtrsim 10-15\msun$) form black holes
through direct collapse and are subject neither to pulsation pair-instability
mass-loss nor to pair-instability supernova disruption for masses 
$M_{\rm He}<90\msun$. Pair-instability disruptions affect only the lowest 
metallicity stars ($Z\lesssim0.0001$) and the most massive stars 
($M_{\rm ZAMS}\gtrsim185\msun$) and the pulsations play no role 
in this model.
}
\label{fig.bhmass}
\end{figure}

\section{Calculations}
\label{sec.calc}

We use the population synthesis code {\tt StarTrack}~\citep{Belczynski2008a}. 
We assume standard wind losses for massive stars: O/B star winds~\citep{Vink2001} and LBV winds 
\citep[specific prescriptions for these winds are listed in Sec.~2.2 of][]{Belczynski2010b}. 
We treat the accretion onto compact objects during the Roche lobe overflow (RLOF) and from stellar 
winds using the analytic approximations presented by \cite{King2001} and by \cite{Mondal2020}, 
and limit accretion during the common envelope (CE) phase to $5\%$ of the Bondi rate 
~\citep{MacLeod2017}. We employ the delayed core-collapse supernova (SN) engine in NS/BH mass 
calculation~\citep{Fryer2012} that allows for populating the lower mass gap between NSs and 
BHs~\citep{Belczynski2012a,Zevin2020}. The most updated description of {\tt StarTrack} is given 
by \cite{Belczynski2020b} and the model M30 in this study describes our standard choices of input 
physics. In our study we employ the fallback decreased NS/BH natal kicks with $\sigma=265\kms$, 
we do not allow CE survival for Hertzsprung gap donors (submodels B in our past calculations), 
and we assume a $100\%$ binary fraction and a solar metallicity of $\zsun=0.02$. 

We extend the initial mass function (IMF) to $200\msun$ and we keep the power-law slope for 
massive stars $\alpha=-2.3$ (in the past we have limited IMF to $150\msun$). This is motivated 
by observations of massive stars; notably three stars in LMC (R136a, R136b,R136c:
~\cite{Bestenlehner2020}) and two stars in the Milky Way (WR 102ka and $\eta$ Car:~\cite{Barniske2008,
Hillier2001}) are estimated to have initial masses close to or exceeding $200\msun$.
We have also adopted favorable (in terms of forming massive BHs from stars) model from 
~\cite{Farmer2020} and from ~\cite{Costa2020} that avoids PPSN mass loss for helium core masses: 
$M_{\rm He}<90\msun$, but allows for disruption of stars above this mass threshold. Such a model 
requires that carbon burning rate is decreased by $2$ standard deviations and that there is an 
episode of dredge-up during core-helium burning phase. 

Orignal stellar evolution formulae that we employ in {\tt StarTack} are based on stellar models 
only up to $50\msun$~\citep{Hurley2000}. Our extrapolation to higher masses was checked to give 
reasonable results in terms of He/CO core masses and/or evolutionary tracks on Hertzsprung-Russel 
diagram in comparison with results obtained with detailed evolutionary calculations (e.g., with 
{\tt Geneva} or {\tt MESA} codes; ~\cite{Belczynski2014,Belczynski2017a,Belczynski2020b}). However, 
we note that there is no current consensus on evolution of massive stars and their calculated radii, 
He/CO core masses and luminosities differ from one detailed calculation to the other (e.g., compare 
Tables 5 and 6 in ~\cite{Belczynski2020b}). 

The results of our model are shown in Figure~\ref{fig.bhmass} in which we present the dependence 
of the final helium core mass on the initial-star mass for various metallicities. The final helium 
core mass is a  good approximation of the BH mass for most massive stars in close binaries. 
The most massive stars are expected to directly collapse to BHs~\citep{Fryer1999,Basinger2020}, 
and stars in close binaries are typically stripped of their H-rich envelopes (BH-BH merger 
progenitors in particular; ~\cite{Belczynski2016b}). Down to metallicity of $Z \sim 0.001$ 
BH masses do not exceed $M_{\rm BH} \sim 50\msun$ which is exactly what we were obtaining with 
our previously employed weak mass loss from PPSN based on calculations of~\cite{Leung2019}. 
Only stars with lower metallicity ($Z \sim 0.001-0.0001$) are affected by our modifications 
and are allowed to form BHs with very high  masses $M_{\rm BH} \sim 50-90\msun$. One notes the 
emergence of the upper mass (at adopted $M_{\rm BH}=90\msun$) for the model with $Z=0.0001$ in 
which BHs do not form for initial star mass above $M_{\rm ZAMS}>185\msun$.

We follow the evolution of Population I and II ($Z=0.03-0.0001$) stars with the input physics 
described above until the formation of BH-BH mergers. We estimate the cosmological BH-BH merger 
rate density using redshift-dependent star-formation history and metallicity evolution across 
cosmic time with the standard Planck-based cosmology~\citep{Belczynski2020b}. Note that we may 
be underestimating the amount of low-metallicity stars~\citep{Chruslinska2019b,Chruslinska2020} 
and therefore our merger rates of most massive BH-BH mergers may also be underestimated.

\begin{figure}
\vspace*{0.4cm}
\hspace*{-0.4cm}
\includegraphics[width=0.5\textwidth]{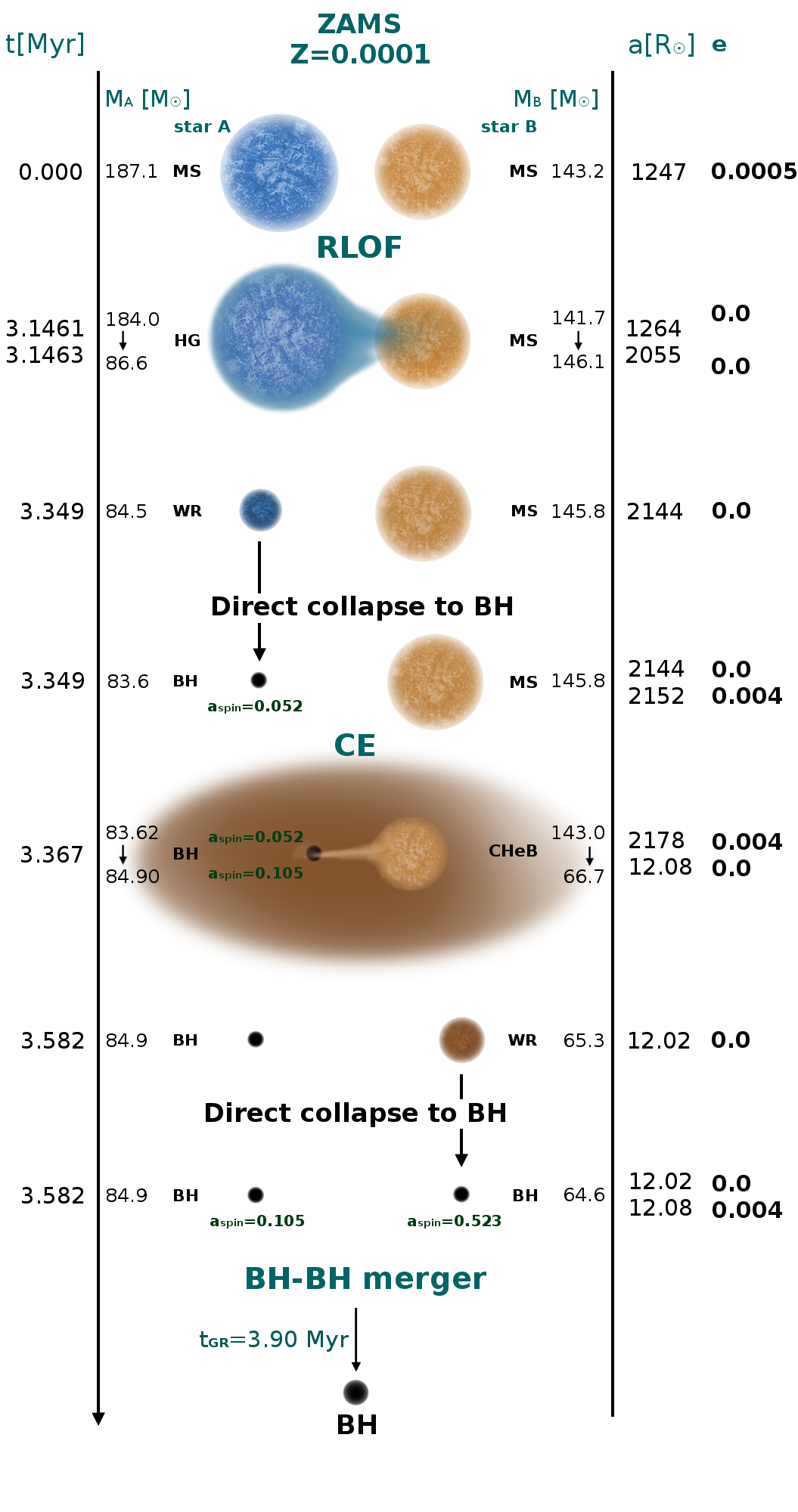}
\caption{
Evolution of an isolated binary system that produces a BH-BH merger resembling GW190521 
at low metallicity ($Z=0.0001$). MS: main sequence star, HG: Hertzsprung gap star, CHeB: 
core helium burning star, WR: Wolf-Rayet star, BH: black hole, RLOF: Roche lobe overflow, 
CE: common envelope.
}
\label{fig.evol1}
\end{figure}

\section{Example of GW190521 Formation}
\label{sec.evol}

In Figure~\ref{fig.evol1} we show an example of evolution: the formation of BH-BH merger 
similar to GW190521 with BH masses $m_1=84.9\msun$ and $m_2=64.6\msun$. The evolution starts 
with a massive primary ($M_{\rm ZAMS,A}=187.1\msun$) and a lighter secondary 
($M_{\rm ZAMS,B}=143.2\msun$) at very low metallicity $Z=0.0001$ on a wide (semi-major axis 
of $a=1247\rsun$) and virtually circular orbit ($e=0.0005$). The primary star evolves off 
the main sequence and becomes a Hertzsprung gap star expanding and initiating a stable RLOF 
that increases the orbital separation ($a=2055\rsun$) and strips the primary
of its H-rich envelope. The primary becomes a massive Wolf-Rayet star ($M_{\rm A}=84.5\msun$)
that soon collapses directly into a BH with mass $M_{\rm BH,1}=83.6\msun$ (no natal kick, 
$0.9\msun$ mass loss in neutrinos) while the secondary is still a main sequence star. When the
secondary becomes a core-helium burning star it expands over its Roche lobe and initiates a CE 
episode. After the CE phase the orbital separation is greatly reduced ($a=12.08\rsun$), 
the primary BH increases its mass through accretion in the CE ($M_{\rm BH,1}=84.9\msun$), and the
secondary loses its H-rich envelope and becomes a massive Wolf-Rayet star 
($M_{\rm B}=65.3\msun$). Then the secondary star undergoes a core-collapse and forms  
directly a second massive BH ($M_{\rm BH,2}=64.6\msun$, no natal kick). Neutrino 
emission induces very a small eccentricity on the BH-BH binary ($e=0.004$) and slightly expands 
the orbit ($a=12.08\rsun$). This BH-BH system has formed after $3.6$Myr of stellar evolution 
and it takes another $3.9$Myr for the two BHs to merge due to emission of gravitational 
radiation and associated orbital angular momentum loss. Due to the very 
short evolutionary and gravitational-wave emission timescale this system would form and 
merge near the redshift it has been detected ($z=0.83$). 

We use the Tayler-Spruit magnetic dynamo angular momentum transport~\citep{Spruit2002} to 
calculate the natal spin of the primary BH ($a_{\rm spin,1}=0.052$: see eq.4 of  
~\cite{Belczynski2020b}). This spin increases due to accretion of $1.3\msun$ in the CE 
($a_{\rm spin,1}=0.105$:~\citep{MacLeod2017}. The spin of the secondary BH ($a_{\rm spin,2}=0.523$: 
see eq.15 of~\cite{Belczynski2020b}) is set by the tidal spin-up of the secondary star when it is 
a compact Wolf-Rayet star in very close binary ($a=12\rsun$ and orbital period of 
$P_{\rm orb}=10$h). Since the BHs were formed without natal kicks we assume that their spins 
are aligned with the binary angular momentum vector. This allows us to assess the effective 
spin parameter of this system: $\chi_{\rm eff}=0.29$, which is within $90\%$ credible 
limits of the LVC estimate ($\chi_{\rm eff}=[-0.28:0.35]$). If the tidal spin-up were not at work 
as envisioned, we would calculate the secondary BH natal spin from our stellar models: 
$a_{\rm spin,2}=0.070$ and that would have resulted in $\chi_{\rm eff}=0.090$. 

In our adopted model massive BHs form through direct collapse of the entire progenitor star into a
BH. Since there is no mass loss we assume no natal kick and the system not only survives the BH 
formation, but also remains aligned (i.e., BH spins are aligned with binary angular 
momentum vector). This leads to an effective precession spin parameter equal zero ($\chi_{\rm p}=0$) 
as precession requires some level of misalignment. This is apparently inconsistent with LIGO/Virgo 
estimate ($\chi_{\rm p}=[0.31:0.93]$), but this estimate is very weak~\citep{gw190521a}. 
Misalignment may be possibly obtained by natal kicks associated with asymmetric neutrino 
emission~\citep{Fryer2006b,Socrates2005} even if there is no baryonic mass ejection at the BH
formation.

\begin{table}
\caption{Merger Rate Densities$^{a}$ [$\gpy$]}
\begin{tabular}{rccccc}
\hline\hline
type            &   $z<0.1$ &    $z<0.4$ &    $z<0.7$ &       $z<1$ &     $z<1.5$ \\
\hline\hline
all NS-NS:      & {\bf 132} &        168 &        203 &         233 &         263 \\
all BH-NS:      &      7.50 & {\bf 11.8} &       17.0 &        22.4 &        31.7 \\
                &           &            &            &             &             \\ 
light BH-BH: &      30.3 &       44.6 & {\bf 63.2} &        84.8 &         131 \\
mixed BH-BH: &     0.028 &      0.055 &      0.115 & {\bf 0.151} &       0.238 \\
heavy BH-BH:    &     0.003 &      0.009 &      0.018 &       0.025 & {\bf 0.038} \\
\hline
\hline
\end{tabular}
$^{a}$: in bold we mark the rate that approximately corresponds to detection horizon 
of a given merger type.
\label{tab.rates}
\end{table}

\section{Populations of BH-BH mergers}
\label{sec.q}

We subdivide the population of BH-BH mergers into three categories: light mergers with 
both BHs having mass $M_{\rm BH}<50\msun$, mixed-mass mergers with one BH with mass 
$M_{\rm BH}<50\msun$ and another with mass $M_{\rm BH}>50\msun$, and heavy mergers with 
both BHs having mass $M_{\rm BH}>50\msun$. The $M_{\rm BH}\approx50\msun$ represents the 
believed (old/outdated) limit for stellar-origin BH formation set by PPSN/PSN. In 
Table~\ref{tab.rates} we present the merger rates of BH-BH subpopulations for a volume 
corresponding to redshift cuts: $z=0.1$ (approximate LIGO/Virgo NS-NS detection horizon),  
$z=0.4$ (BH-NS horizon), $z=0.7$ (light BH-BH horizon), $z=1.0$ (mixed-mass BH-BH horizon), 
$z=1.5$ (heavy BH-BH horizon).

Our merger-rate estimates are consistent with the $90\%$ LVC~\citep{LIGO2019b} empirical 
estimates: for NS-NS we find $132\gpy$ (LVC O1/O2: $110$--$3840\gpy$), BH-NS $11.8\gpy$ 
(LVC O1/O2: $<610\gpy$), light BH-BH $63.2\gpy$ (LVC O1/O2: $9.7-101\gpy$). For heavy 
BH-BH mergers we find a rate of $\sim 0.04\gpy$ (LVC O3: $0.02-0.43\gpy$ rate based on the
single detection of GW190521). This may seem to be a marginal match but note that the LVC 
estimates are only $90\%$ credible limits. Merger rates are subject to change with 
various assumptions about input physics (natal kicks, CE, cosmic evolution of metallicity:  
~\cite{Belczynski2020b}) and they will be re-evaluated once the LVC provides more 
restrictive estimates.

\begin{figure}
\vspace*{0.4cm}
\hspace*{-0.4cm}
\includegraphics[width=0.5\textwidth]{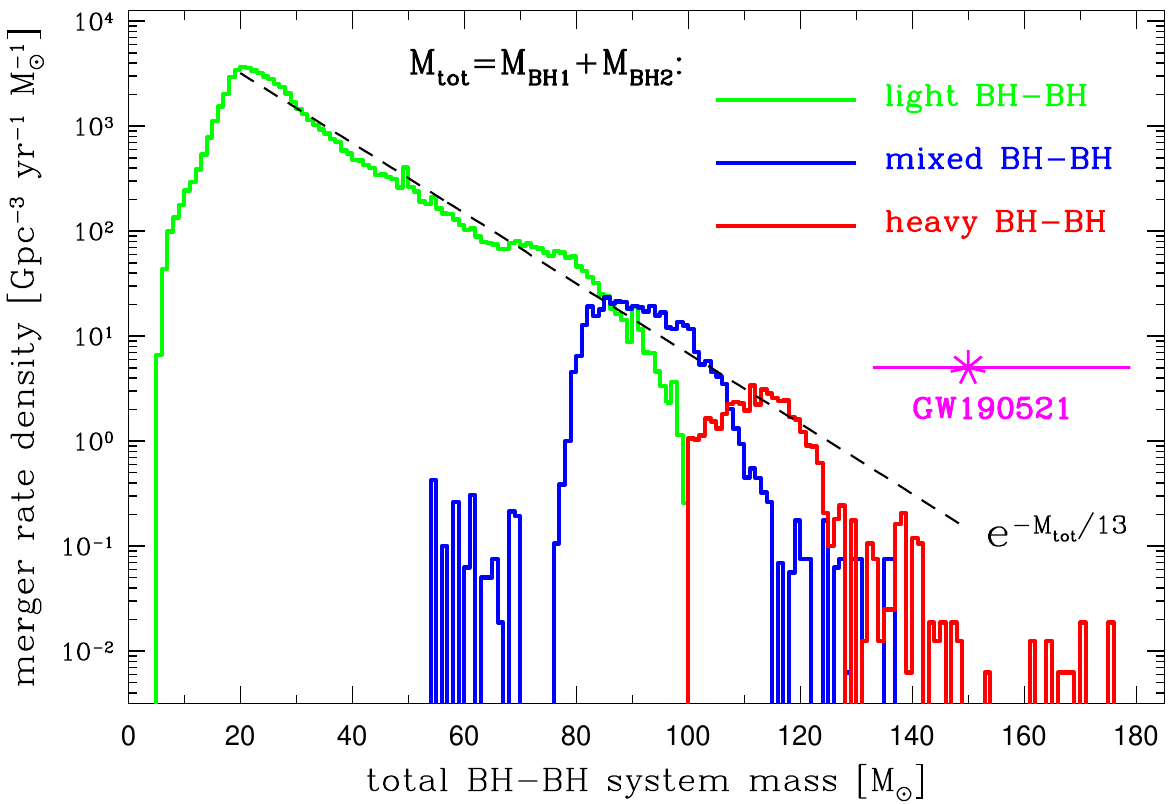}
\caption{
Total intrinsic mass distribution for the three subpopulations of BH-BH mergers ($z<1$).
Note that GW190521 is found in the tail of distribution of heavy BH-BH mergers.  
}
\label{fig.mtot}
\end{figure}

\begin{figure}
\vspace*{0.4cm}
\hspace*{-0.4cm}
\includegraphics[width=0.5\textwidth]{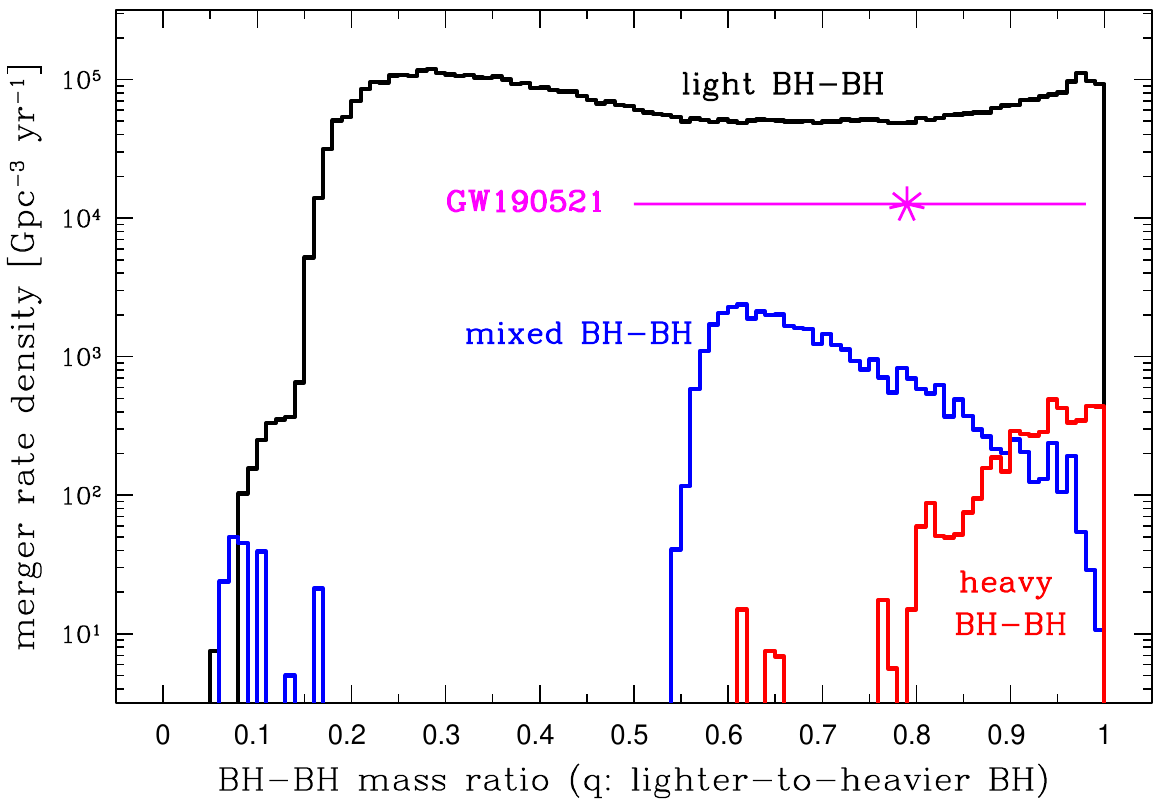}
\caption{
Intrinsic mass ratio distribution for the three subpopulations of BH-BH mergers 
($z<1$). 
}
\label{fig.q}
\end{figure}

In Figure~\ref{fig.mtot} we show the intrinsic (not redshifted) distribution of the total 
BH-BH binary mass for mergers found in the redshift range $z<1$. By construction, the 
light BH-BH mergers are found with $M_{\rm tot}=5-100\msun$, where the lowest masses  are 
reached for $\sim 2.5+2.5$ mergers with both BHs originating from our delayed SN engine
~\citep{Fryer2012,Belczynski2012a} and thus allowed in the lower ``mass-gap'', while the 
heaviest $\sim 50+50$ mergers form with PPSN mass loss~\citep{Leung2019}. The heavy mergers 
have total mass in the range $M_{\rm tot}=100-180\msun$, although the number of BH-BH 
mergers rapidly declines with increasing mass. This comes from the assumption that the IMF 
is steep (power-law with exponent $-2.3$) for massive stars. In fact, the overall 
population of BH-BH mergers show a rapid decline of number of mergers with mass from light 
systems to mixed (intermediate-mass) systems to heavy systems. Note that the total BH-BH 
binary mass declines like an exponential (evolutionary processes affecting IMF) and not 
like a power-law that is commonly assumed in literature. GW190521 with a total mass 
of $M_{\rm tot}=150^{+29}_{-17}\msun$~\citep{gw190521a} is found in the tail of the mass 
distribution of our heavy BH-BH mergers. If future observations will show a flatter BH-BH 
mass spectrum, it would be an indication that some evolutionary process must be at work. 
For example, in our model the natal kicks operate only for the lightest BHs 
($M_{\rm BH}\lesssim10-15\msun$) and are decreasing with BH mass creating a peak in total 
BH mass at $M_{\rm BH}\sim20\msun$. Had we allowed natal kicks to be applied differently 
it would be possible to flatten the BH mass spectrum in a desired mass range and possibly 
place some constraints on the core-collapse asymmetries. 

In Figure~\ref{fig.q} we show the intrinsic mass ratio ($q=M_{\rm BH,2}/M_{\rm BH,1}$ with 
$M_{\rm BH,1} \geq M_{\rm BH,2}$) distribution of BH-BH mergers found in redshift range $z<1$.
The light BH-BH mergers show rather flat mass ratio distribution in a broad range $q=0.2-1$ 
and tail reaching down to $q=0.05$, with two small peaks: one at $q \sim 0.25$ and another 
at $q \sim 0.95$. The latter peak is a standard result of isolated binary evolution when 
rapid SN engine (that does not produce BHs in the lower mass gap: $M_{\rm BH}<5\msun$) is 
applied to calculate BH mass and BH-BH mergers with similar mass BHs dominate the population 
(e.g.,~\cite{Belczynski2016b}). However, note that BH-BH mergers can still reach mass ratios 
as small as $q \sim 0.2$~\citep{Olejak2020b}. The former peak, and the extent of mass ratio 
to very small values, is the result of our application of the delayed SN engine to calculate 
BH masses and our assumption that the NS/BH mass limit is at $2.5\msun$. The population of 
relatively abundant (IMF) low-mass BHs (e.g., these in the lower mass gap: 
$M_{\rm BH}\sim 2.5-5\msun$) forms in binaries with more massive BHs creating the low-$q$ 
BH-BH mergers.The lowest mass ratio arises from extreme systems with $2.5+50\msun$ BH-BH 
mergers. Even more extreme mass ratio systems are found in BH-NS merger populations
~\citep{Drozda2020}. 

The heavy BH-BH mergers are limited to $q \gtrsim 0.6$ as the lowest mass BH in this 
subpopulation is $50\msun$ and the heaviest $90\msun$. Since this subpopulation does not 
include low-mass BHs it tends to produce similar component mass BH-BH mergers with typical 
mass ratio of $q \sim 0.9-1$. This is consistent with LVC estimate of GW190521 mass ratio 
$q=0.79^{+0.19}_{-0.29}$~\cite{gw190521a}.

\section{Conclusions}
\label{sec:concl}
 
We extended our evolutionary model to stars up to $200\msun$ and we limited the action 
of mass loss associated with pair instabilities~\citep{Farmer2020} to test whether it is 
possible to form BH-BH mergers resembling GW190521 that hosts $85\msun$ BH and $66\msun$ 
BHs through classical isolated-binary evolution. Such massive BHs were/are believed not to 
form directly from stars. 

It is in fact possible to form massive BHs in BH-BH mergers resembling GW190521 if 
C-burning reaction rate uncertainties that may limit the pair-instability associated mass 
loss are taken into account. Once such possibility is adopted, our standard binary
evolution delivers merger rates of ``normal'' BHs (light BHs: $<50\msun$) and heavy BHs 
($>50\msun$) that are consistent with LIGO/Virgo observations. 

The binary evolution leading to the formation of systems resembling GW190521 is relatively 
simple. It requires two very massive stars ($M_{\rm ZAMS}\sim 150-200\msun $) at low 
metallicity ($Z \sim 10^{-4}$) and it involves a stable RLOF and CE episode. Our standard 
assumptions on BH formation involves direct BH formation through standard core-collapse 
for both BHs with no associated PPSN mass loss and with no natal kicks. 

The binary evolution leading to the formation of GW190521-like mergers may or may not involve 
tidal spin-up of WR stars that are the immediate progenitors of massive BHs. In both cases 
the low predicted effective spin parameter of our proposed BH-BH 
merger example ($\chi_{\rm eff}=[0.09:0.29]$) is consistent with LIGO/Virgo observations
($\chi_{\rm eff}=[-0.28:0.35]$). In either case, the measurement of GW190521 effective spin 
is consistent with efficient angular momentum transport in massive stars by a magnetic dynamo. 

Our model predicts that effective precession spin parameter (measuring misalignment of BH 
spins from binary angular momentum) for GW190521-like systems is negligible $\chi_{\rm p}=0$.
This is inconsistent with the LIGO/Virgo estimate: $\chi_{\rm p}=[0.31:0.93]$. However, this 
empirical estimate was exposed as highly uncertain and a non-precessing interpretation of 
GW190521 cannot be excluded~\citep{gw190521a}. If precession is confirmed in such mergers 
it either indicates that they do not form through a classical isolated binary evolution 
channel or that the second BH formation is asymmetric and leads to non-negligible BH natal 
kick (misalignment). 

Finally, we emphasize that these new results are only valid if the carbon fusion reaction 
rate is highly uncertain and is allowed to be $\sim 2$ standard deviations below the 
standard {\tt STARLIB} rate, which is unlikely but not impossible~\citep{Farmer2020} and 
if during core-helium burning phase there is an episode of a dredge-up~\citep{Costa2020}.

\vspace*{-0.2cm} \acknowledgements
KB acknowledges support from the Polish National Science Center grant Maestro 
(2018/30/A/ST9/00050). Acknowledgment of support/advice goes to A.Olejak, P.Drozda, 
V.Babihav, L.Oskinova, J.-P.Lasota, R.Farmer, T.Bulik, I.Mandel, S.Woosley, M.Giersz, 
R.O'Shaughnessy, M.Chruslinska, J.Klencki.

\bibliography{biblio}

\end{document}